\documentstyle[preprint,aps]{revtex}
\tightenlines
\begin{document}

\draft

\title{Pre-equilibrium evolution of quark-gluon plasma}

\author{Gouranga C Nayak \thanks{ e-mail:gcn@iitk.ernet.in}
and V. Ravishankar \thanks{e-mail:vravi@iitk.ernet.in}}
\address{Department of Physics, Indian Institute of 
technology, Kanpur -- 208 016, INDIA }

\maketitle

\begin{abstract}
We study the production and the evolution of quark-gluon plasma
expected to be formed in ultra relativistic heavy-ion
collisions, within the color flux-tube model. We introduce 
the gluonic component in the Boltzmann equation, which 
we solve in the phase space which is extended to include
the SU(3) color degree of freedom. The color degree of 
freedom is shown to play a decisive role in equilibration,
and in fixing the temperature of the plasma. We further 
find that the soft partons that we study here contribute
substantially to the bulk properties. Finally, the model
is shown to provide a detailed picture of how the 
quark-gluon plasma evolves and is driven towards a 
hydrodynamic flow, which starts occuring around 1 $fm$.

\end{abstract}

\vspace{1.5 cm}

\section{INTRODUCTION}
It is by now well established from lattice studies
\cite{lattice} that hadronic matter
at sufficiently high temperatures ($\sim 200 MeV$) and
pressures undergoes a transition to the so called
quark gluon phase involving, i) deconfinement
and ii) chiral symmetry restoration, not necesarily
simultaneously.
It also appears that the deconfinement transition 
is of the first order, and that the deconfined phase
near the critical temperature is still rather far
away from showing an ideal gas behaviour. In any case,
while such a phase surely did exist in the early
universe, it is interesting that we may expect to
produce this phase in ultra relativistic heavy ion
collisions (URHIC). In fact it is widely believed that
RHIC and LHC will succeed in revealing 
this new phase of matter.

The deconfined phase is not accessible directly in 
the accelerator experiments. The signatures are therefore
necessarily indirect,
and the prominent ones that have been studied are 
: 1) $J/{\psi}$ suppression \cite{satz},
2) electromagnetic probes such as dilepton
and direct photon production
\cite{strickland,sinha}, and 3) strangeness
enhancement \cite{singh}.
However, a proper diagnostics for these 
signatures in the accelerator produced plasma 
must involve a study of the evolution of the system
in all its stages, including its production soon after
the two nuclei have collided with
each other, the equilibration, hydrodynamic expansion
leading to cooling, and finally hadronization. Note that
the above stages
are not necessarily mutually exclusive; 
there could be an overlap
between production and equilibration, as also between cooling
and hadronization. 
Of particular importance to us here are the production and the
equilibration regimes; indeed it has been pointed
out\cite{satz,satz1} that $J/{\psi}$ suppression
which is generally attributed to the equilibrium
stage can also arise from the pre-equilibrium stage,
an observation which is true for strangeness enhancement
as well. 
The contribution of the pre-equilibrium stage to 
the dilepton production and the direct photon
production would be
even more pronounced, and a careful study of the above 
signals ought
to shed light on this stage of the plasma.

In this paper we discuss the production and the evolution of
quark-gluon plasma(QGP) in URHIC by employing the
Flux-tube Model \cite{baym,kajantie}
which is a generalization of the familiar Lund string model
widely used for $e^{+}e^{-}$ and $p-p$ collisions 
\cite{andersson}.
This model allows for
a concurrent production and evolution of quark pairs and gluons
by a background chromoelectric field.
Assuming the Bjorken scenario \cite{bjorken} where a 
plateau will be seen in the central
rapidity region in 
relativistic heavy-ion collisions,
the Flux-tube model deals with the 
baryon free plasma in the 
central region; there would initially be a huge deposit 
of energy, modeled by the creation of a 
chromo field between the two receding
nuclei, which decays spontaneously to produce partons.
Note that the field configuration which contains this energy is
necessarily electric like if it has to produce partons.
Collisions between the partons 
and their acceleration ( which is again due 
to the background field)
 dictate the dynamics of their evolution and equilibration.
We propose to study this dynamics
within the classical Boltzmann
equation, by explicitly incorporating the production in 
the source term, acceleration by 
the background chromo-field and also writing
down a collision term.
	   
In a recent paper \cite{nayak} we studied the
evolution of a non-abelian q$\bar{q}$ plasma in the context of 
color flux-tube model, where it was shown that the non-abelian
features have a major role in the evolution of the system in
a manner that was not captured in any of the earlier studies
\cite{baym,kajantie,banerjee}
which were purely abelian in their content.
It was also argued that it was unlikely that the system
would equilibrate instantaneously. 
However, for simplicity, we had 
ignored there the gluonic component
and considered the simpler gauge group SU(2).
In this paper, we remedy both the drawbacks; we consider the
gauge group SU(3) as is appropriate for a real plasma. We also
include the gluons, which have been completely ignored so
far.
We pay attention to the detailed evolution, and its 
prediction for different bulk 
properties of the plasma in this paper.
Signatures such as dilepton, direct photon and strangeness
production from
this model will be calculated and reported
elsewhere.

There is yet another class of models which describe
the evolution of quark-gluon plasma such as 
heavy-ion jet interaction generator(HIJING) \cite{hijing} and
parton cascade model(PCM) \cite{pcm}. 
These models are governed by perturbative QCD (pQCD).
The production and the evolution of hard and semi-hard partons
are studied by a master rate equation in HIJING 
and by a transport equation in PCM. 
While these may be reliable in the study 
of hard partons, these perturbative
approaches are admittedly insufficient to 
study the dynamics of the
soft partons \cite{hijing}, in particular, their production. 
Being perturbative, they do not incorporate any of the 
non-perturbative
aspects such as the formation of the strings
and their break up,
which is studied even in $pp$ collisions. 
The addition of soft partons to hard and
semi- hard partons 
changes the bulk properties of plasma, such as temperature and 
energy density. Xu {\it et al.} \cite{satz1} have observed
that this addition leads to an enhanced suppression of 
$J/{\psi}$, which can be understood to be a consequence of
increased number density and a lowered temperature of QGP.
In our study, we employ the Schwinger mechanism 
\cite{schwinger}
for particle production, which is quintessentially 
non-perturbative.
Even if one were to employ a perturbative version, 
as a time dependent
electric field would require \cite{bhal2},
 it may be noted that an initial electric
field as a classical saddle point owes its existence 
to non-perturbative
 processes, {\it viz}, the soft gluon 
 exchanges that take place between
the two nuclei.
On the whole, one may expect that the pQCD based studies
\cite{hijing,pcm}
will be useful in the hard regime, and that the 
flux tube like models will be
required to study the soft regime which 
will become increasingly
prominent as the system expands and more and more
secondaries are produced \cite{scpcm}. 
Indeed, it should be possible to develop a unified approach
to study the hard and the soft components, say  {\it a la} the
approach of Eskola and Gyulassy \cite{eskola} who have
included the minijets in their so called chromoviscous
hydrodynamics which is again based on the flux tube model.
This study will be taken up separately.

The paper is organised as follows.
In section {\bf II} we set up the transport 
equations for the quarks and the gluons 
in an extended phase space. Section {\bf III} contains a
description of the color flux-tube
model, and its incorporation into the transport equations. 
The numerical procedure is presented in section {\bf IV} and 
the results 
are discussed and compared with other
models in section {\bf V}.
We conclude and summarize the main results in section {\bf VI}.

\section{Transport equation in extended phase space}
 
 In non-abelian theory the color charge is a continuously
 varying function of time. 
 The precession of the color charge($Q^a$)
 obeys Wong's equation \cite{wong}:

\begin{equation}
{dQ^a\over d{\tau}}= f^{abc} u_\mu Q^b A^{c\mu} 
\label{one}
\end{equation}

 which supplements the Lorentz force equation

 \begin{equation}
 {dp^{\mu} \over d{\tau}} = Q^a F^{a\mu\nu}u_{\nu}.
 \end{equation}

Here $A^{a\mu}$ is the gauge potential, and
$f^{abc}$ is the structure constant of the gauge group.
In order to include the color charge in the
phase-space,  we consider an extended
one particle phase space of dimension $d = 6 + (N^2 - 1)$,
(with $N=3$). The extended phase space is taken to be the 
direct sum $R^6 \oplus G$, where $G$ is the(compact) space 
corresopnding
to the given gauge group. 
In short, in addition to the usual 6 dimensional 
phase space of coordinates and
momenta we now have another eight  coordinates corresponding
to the eight color charges in SU(3). 
In this extended phase space a typical transport
equation reads as \cite{heinz}:

\begin{equation}
\left[ p_{\mu} \partial^\mu + Q^a F_{\mu\nu}^a p^\nu 
\partial^\mu_p
+ f^{abc} Q^a A^b_\mu
p^{\mu} \partial_Q^c \right]  f(x,p,Q)=C(x,p,Q)+S(x,p,Q)
\end{equation}

Here $f(x,p,Q)$ is the single particle distribution function
in the extended  phase space. The first term in the lhs of 
equation (3) corresponds
to the usual convective flow, the 
second term is the non-abelian version of
the Lorentz force term and the 
last term corresponds to the precession
of the charge as described by Wong's equation.
$S$ and $C$ on the right hand side of equation (3) 
correspond to the source and collision terms respectively
(described below).
Note that we have to write
separate equations for quarks, antiquarks and gluons since they
belong to different representations of $SU(3)$.

The term gluonic source merits some explanation here. The classical
background field that we consider here has, in contrast to the Maxwell
field, self interaction. We are interested in the stability of the
gluonic vacuum ( which is the analogue of the 
radiation in eletrodynamics)
against the fluctuations in the classical background field. 
An adaptation
of the Schwinger mechanism in QED shows that the fluctuations can indeed
produce the gluons, {\it i.e.} the off-shell classical field can 
spontaneously
produce the on-shell radiative gluonic field (see equation
9). There is, therefore, no
ambiguity or double counting in this process. 
The source term yields
asymptotic gluonic states, and further
interaction between the
gluons is treated separately by a collision term.

\section{The Flux-tube Model}
We briefly review the flux tube model as is appropriate to
the context here.  
In this model, the two nuclei that undergo a
central collision at
very high energies are Lorentz-contracted as thin plates. 
When these two Lorentz-contracted 
nuclei pass through each other
they  acquire a nonzero color
charge ($<Q> = 0, <Q^2> \neq 0$), by 
a random exchange of soft gluons.
The nuclei which act as color 
capacitor plates produce a chromo-electric field between them
\cite{low,nussinov}. The strength of 
the field which naturally depends
on the strength of the color charge residing on the plates
cannot be fixed from first principles. 
We can only fix that phenomenologically, say by identifying
the field energy with the energy in the 
central region as estimated by Bjorken \cite{bjorken}.
This strong electric field creates $q\bar{q}$ and gluon pairs
via the Schwinger mechanism which 
enforces the instability of the
vacuum in the presence of an external field. The partons so
produced, collide with each other and also get accelerated
by the background field. As described in the previous section,
these color charges rotate in the color space, a feature
which is manifest in this model, 
but largely ignored in the earlier
studies \cite{baym,kajantie,banerjee}.
Thus the production, collision, acceleration and
rotation are implemented in one single transport equation,
i.e. equation (3).

 It is very difficult to solve the transport equation written
 above, in general.
And we have a set of three coupled equations here.
As was done in reference \cite{nayak} 
we make a few assumptions.
First of all, we admit only those potentials 
which can be brought to a form where 
the only surviving components are
$A^{{\mu}a} = (A^{01}, A^{31})$. This
choice restricts $F^{{\mu} {\nu}}$ to be ``Maxwell" like, 
also pointing  in the ``1" direction in the color space. 
This restriction is not
arbitrary as it is known \cite {brown} that the 
non-Maxwellian configurations - where
the charges, the gauge potential
and the fields do not lie in the same 
direction in the group space -  do
not produce particle pairs, in general.
Secondly, we require a boost invariant 
description of the physical
quantities \cite{bjorken}. Accordingly, we demand that the
distribution functions also be functions 
of only boost invariant quantities. 
Finally, we work within the Lorentz
gauge  which is implemented elegantly by the choice
$ A^{\mu a}=\epsilon^{\mu\nu} \partial_{\nu} G^a(\tau),
~~ \mu, ~ \nu =0,3$, with all the other components zero.
$\tau = (t^2 - z^2)^{1/2}$ 
is the boost invariant proper time.

  Now we fix the magnitude of the vector charge $Q^a$, 
which corresponds to the first Casimir invariant of 
SU(3). It is simply the value of the coupling
constant. There is also another Casimir invariant, {\it viz},
$ d^{abc}{Q^a Q^b Q^c}$, which is also conserved as the QGP
evolves. There is, however, no way of fixing its value and 
the experiments presumably impose no restriction on its allowed
values. In fact, the same holds for lattice analyses as well.
For this reason, we do not take cognizance of this invariant,
and we conveniently resolve the SU(3)
charges in the polar coordinates: 
$Q_i = Q \prod_{k=1}^{i-1} sin\theta_k
cos\theta_i$ for $i \neq 8$,
and $Q_8 = Q \prod_{k=1}^7 sin\theta_k$.

We now fix the collision term. A collision 
term can be indeed obtained
from pQCD. Apart from making the 
equation hopelessly non-linear,
this choice would be good only for the hard components, which
are not of interest to us here. 
As aptly pointed by Hung
and Shuryak \cite{shuryak} recently, a microscopic 
description of collisions
gets increasingly cumbersome and also unnecessary as more and
more secondaries are produced. On the other 
hand it is admittedly
true that there is no way to reliably obtain a collision term
in the non-perturbative regime. So we shall employ a relaxation
time approach here, where the relaxation 
time $\tau_c$ will have to
be fixed phenomenologically, and in all probability, {\it a 
posteriori}. Recently it is proposed \cite{Wong,heiselberg} that 
$\tau_c$ can be local and have a (weak) 
dependence on $\tau$. We
take $\tau_c$ to be constant here,
and further refinements may be 
incorporated after we have a better 
understanding of the transport
phenomena both experimentally and theoretically.

Within the relaxation approach, the collision 
term is written as
\begin{equation}
C = {-{p^\mu u_\mu (f-f^{eq})} \over {\tau_c}}.
\end{equation}
Here $f^{eq}$ is the distribution function,
in local equilibrium. Note that the locality can extend
to the color space as well apart from space-time.
Taking it to be an ideal gas, for simplicity,
we write, 
\begin{equation}
f_{q,g}^{eq} = \frac{2}{\exp ((p^\mu-Q^a A^{\mu a})
u_\mu /T(\tau)) \pm 1}
\end{equation}
where the +(-) sign is to be taken
for fermions (bosons). Note that
we allow a common equilibrium temperature 
for quarks and gluons.
Here $u^{\mu} = (cosh\eta,0,0,sinh\eta)$ 
is the flow velocity and 
$\eta$ is the space time rapidity given by $tanh\eta = z/t$.
With our choice of potentials, it follows that  
$A^{\mu} u_{\mu} = 0$, so that the equilibrium distribution 
function can be written as:
\begin{equation}
f_{q,g}^{eq} = \frac{2}{\exp ((p^\mu u_\mu) /T(\tau)) \pm 1}
\end{equation}

Note also that the temperature depends only on the proper time,
in accordance with the Bjorken picture. 
Demanding the same of the
distribution functions as well, we require that
the longitudinal boosts be symmetry operations on the 
single particle distribution. The boost-invariant 
parameters are, apart from the color charges,
\begin{equation}
\tau =(t^2-z^2)^{1/2},\, \xi=(\eta-y) ,\, 
p_t=(p_0^2-p_l^2)^{1/2}
\end{equation}
where $y=\tanh^{-1}(p_l/p_0)$ is the momentum rapidity.

The above set of invariant variables also serve to express the
source terms for $q\bar{q}$ and gluon pairs which
will be obtained by the Schwinger mechanism.
The expression for 
$q \bar{q}$ production (obtained from constant electric field)
is written as \cite{nayak}
\begin{equation}
S_q(\tau,\xi,p_t,\theta_1)=-\frac{gE
\cos\theta_1}{8\pi^3} 
\ln \left[ 1-\exp \left(
-{2\pi p_t^2 \over gE \cos\theta_1} \right) \right] 
({\alpha \over \pi})^{1/2}
\exp (-\alpha\xi^2)
\end{equation}
and for gluon pair production the corresponding term in SU(3)
is given by the relatively stronger term \cite{heinz,gyulassy}
\begin{equation}
S_g(\tau,\xi,p_t,\theta_1)= (3/2)
S_q(\tau,\xi,p_t,\theta_1).
\end{equation}

We put $g=4$ throughout our calculation.

Now consider the third term 
$f^{abc} A^{\mu a} Q^b {\partial \over 
{\partial {Q^c}}} f(x,p,Q)$,
in the transport equation (3).
Due to the restriction of the gauge
potentials to the form
$A^{\mu a} = (A^{0 1},A^{31})$,
there is an additional simplification
in the transport equation. In order to see that, note that
equation (3) has the formal solution

\begin{equation}
f_{q,g}(\tau,\xi,p_t,Q)=\int^\tau_0 d \tau^\prime 
\exp (\frac{\tau^\prime-
\tau}{\tau_c}) \left[ \frac{S_{q,g}
(\tau^\prime,\xi^\prime,p_t,Q )}
{p_t \cosh \xi^\prime} + \frac{f_{q,g}^{eq}
(\tau^\prime,\xi^\prime,p_t,Q)}
{\tau_c} \right],
\end{equation}
where $\xi^{\prime}(\tau\prime)$ is given by

\begin{equation}
\xi^\prime= \sinh^{-1} \left[ \frac{\tau}{\tau^\prime}
\sinh \xi
+ \frac{g\cos\theta_1}{p_t \tau^\prime} 
\int_{\tau^\prime}^ {\tau} 
d\tau^{\prime\prime} E(\tau^{\prime\prime}) \right].
\end{equation}
Clearly, 
$f^{1bc} A^1 Q^b {\partial \over {\partial Q^c}} f(x,p,Q)
= 0$, as S, $f_{eq}$ and $\xi^{\prime}$ 
only depend on $Q_1$. 

With this final simplification, 
 we get a set of three equations, one each for quark,
antiquark and gluon respectively. They are, explicitly,
 \begin{eqnarray}
\left[ {\partial  \over \partial \tau}
-\left( {\tanh \xi \over \tau}  + {g \cos \theta_1 
E(\tau) \over p_t \cosh \xi} \right) 
{\partial  \over \partial\xi}\right] \nonumber
f_{q,g}(\tau,\xi,p_t,\theta_1
) \\
+ {f_{q,g} \over \tau_c} = {f_{q,g}^{eq} \over \tau_c}
+ {S_{q,g} (\tau,
p_t,\xi,\theta_1) \over p_t \cosh \xi}
\end{eqnarray}

for quarks and gluons, and  

 \begin{eqnarray}
\left[ {\partial  \over \partial \tau}
-\left( {\tanh \xi \over \tau}  - {g \cos \theta_1 
E(\tau) \over p_t \cosh \xi} \right) 
{\partial  \over \partial\xi}\right] \nonumber
{\bar f}_q(\tau,\xi,p_t,\theta_1
) \\
+ {{\bar f}_q \over \tau_c} = {f_q^{eq} \over \tau_c}
+ {S_q (\tau,
p_t,\xi,\theta_1) \over p_t \cosh \xi}
\end{eqnarray}
for antiquarks.

In the process where the field and the 
particles are present the conservation
of energy-momentum is expressed in the form: 
\begin{equation}
\partial_\mu T^{\mu\nu}_{mat} + 
\partial_\mu T^{\mu\nu}_{f}=0,
\end{equation}
where
\begin{equation}
T^{\mu\nu}_{mat}=\int p^\mu p^\nu 
(2 f_q + 2 \bar{f_q} + f_g)d\Gamma d {\Omega_7},
\end{equation}
and
\begin {equation}
T^{\mu\nu}_f = \mbox{diag} (E^2/2 ,E^2/2 ,E^2/2 ,-E^2/2 ).
\end{equation}
Here $d \Gamma = d^3 p/{(2 \pi)^3 p_0} = 
p_t d p_t d \xi/{(2 \pi)^2}$,
and $d \Omega_7$ is the angular integral 
measure in the color space.
The factor 2 in the equation (15) is for two flavors of quarks.
Now since energy and momentum are conserved in each collision, 
we have:
\begin{equation}
\int p^{\nu} C d \Gamma d \Omega_7 = 0.
\end{equation}

Taking the first moment of the Boltzmann equation, integrating
over the color degrees of freedom for
$f_q$, $\bar{f}_q$ and $f_g$ and
making use of equations (14) and (17), 
we obtain from equation (3)

\begin{equation}
\partial_\mu T^{\mu\nu}_f + gE(\tau ) 
\int d\Gamma d {\Omega}_7 p^\nu
\frac{\partial (2 f_q- 2 \bar{f}_q +f_g)}{\partial \xi}  
+ 4 \int d\Gamma d {\Omega}_7 p^{\nu} S_q 
+ 2 \int d\Gamma d {\Omega}_7 p^{\nu} S_g = 0 
\end{equation}
In the above equation the factor 4 in third term 
arises because we have   
two separate transport equations for
quark and antiquark, each coming
with two flavors, and the factor 2
in fourth term is due to $gg$ pair production, 
although there is only one transport equation for gluon.
Putting $\nu$ = 0 and 3 in (18) we get two equations,
which then
yield the following equation for the decay of the electric field:
\begin{equation}
{ dE(\tau) \over d\tau }- \frac{2g\gamma }{2\pi} 
\int^\infty_0 dp_t p_t^2
\int^\infty_0 d\xi\sinh\xi \int^{\pi}_0 d\theta_1
[2 f_q - 2 \bar{f}_q +f_g ] 
+({{\pi}^3/6})\bar{a} \vert E(\tau) \vert^{3/2} = 0.
\end{equation}

Here $\bar{a} = a\zeta (5/2)\exp (0.25/\alpha )$,  
$ a = c(g/2)^{5/2}\frac{7}{2 (2\pi)^3}$ and  
$ c = \frac{2.876}{(4\pi^3)}$.
Finally, $\zeta (5/2) =1.342$ is the Riemann zeta function.

To solve this equation we fix the form of 
$T(\tau)$, by demanding that
the particle energy density differ negligibly 
from the equilibrium 
energy density, in each collision. We 
then relate the proper energy 
density, which is defined by
\begin{equation}
\epsilon(\tau)=\int d\Gamma d {\Omega}_7 (p^\mu u_\mu)^2 
(2 f_q + 2 \bar{f}_q +f_g),
\end{equation}

to the temperature by its equilibrium value, whence,

\begin{equation}
T(\tau)=[{{10 \epsilon(\tau)} \over {\pi^6}}]^{1/4}.
\end{equation}
We solve the equation (19) numerically
to see the evolution of quark-gluon plasma.

\section{Numerical procedure}
The numerical procedure is already 
discussed in earlier works \cite{nayak,banerjee}. We use a 
double self-consistent method.
The procedure follows the scheme $\{T(\tau)_{trial},\, 
E(\tau)_{trial}\}    
\, \, \rightarrow \{f, \bar{f}, \, \, E(\tau)\} \, \, 
\rightarrow
\{f,\bar{f}\} \, \, \rightarrow T(\tau) \, \, \rightarrow $ 
...
by repeated use of equations (10), (19), (10), (21),
which is iterated until there is a convergence to the required
degree of accuracy. 
We have in mind the LHC energies, and
we have taken the initial energy
density as $\epsilon_0 = 300 GeV/{fm^3}$. 
Compared  to PCM \cite{pcm}
where the initial particle energy density 
is taken to be around 
1300 Gev/${fm^3}$
at RHIC, our choice of initial field energy 
density might appear some what low. 
However, our choice is guided by
the estimate that the formation
time for a qgp is a fraction of a fermi, as we explained in ref
\cite{nayak}. In any case it would not be
very appropriate to compare the two initial conditions since
the "initiality" is only in a limited sense. Indeed, it is the
energy density at the location of the receding color plates,
and being well in the fragmentation region at all later times
after the collision, what matters is the energy in the 
central region, corresponding to larger and larger values of
$\tau$ as the system evolves. It will be
seen that in this region
the results that our study yields are not unreasonable, if we
make a judicious choice for the value of $\tau_c$.
We have studied, in this paper, the results 
for three different values of $\tau_c$. For hydrodynamic 
and collisionless limits we have choosen
$\tau_c =$ 0.001 $fm$ and 5.0 $fm$.
We have compared the results of these limitng value
of $\tau_c$ to a realistic intermediate value $\tau_c=0.2 fm$.

\section{RESULTS AND DISCUSSIONS}
The solution of the transport equations following the procedure
outlined in the previous section allows us to
determine directly
the temporal evolution of the particle and energy densities,
temperature and also the rate at which the field energy flows
into the particle sector. These quantities are
of intrinsic interest
and are, in principle, amenable to experimental
study via dilepton
production. We present these results below.
Also of interest are
the broader questions: when the equilibration sets in, at what
time the flow becomes hydrodynamic - with or without viscous 
flow \cite{bjorken,gyulassy}, etc. 

Then there are features peculiar
to our model. Since the color plates are receding
away from each 
other, energy is continuously pumped into the field,
which subsequently
decays to produce particles. We study the 
relative rates at which
these two dynamical processes proceed in URHIC. This process cannot
of course proceed indefinitely. 
This has already been observed by Gyulassy and Csernai in their study
of the dynamics of fragmentation region \cite{csernai}. As the plates give up their
energy, they decelerate. The deceleration sets limits on the times up to
which the model is valid. Indeed, the assumption of boost invariance 
breaks down as it is strictly exact only if $v_{plate} = c$. It is safe
to 
assume boost invariance so long as 
$v_{plate} \geq 0.9 c$ \cite{fnote1}. A 
simple estimate shows that  this condition, 
for our choice of initial energy density,
holds up to $\sim 5-10 fm$ for
LHC energies. There would be other competing processes in the 
fragmentation
region, and the Bjorken scenario is probably valid upto $3-4 fm$. 
Keeping this 
in mind, we have restricted ourselves to (proper) times $\leq 1.5 fm$.
Note that the results presented below will be applicable for 
RHIC energies only if $\tau < 1 fm$.

Finally, the formulation
allows us to study an inherently non-abelian quantity,
{\it viz}, \\
$<cos^2\theta>$, where $\theta$ is the
angle between the charge and the field in the
group space. Note that its abelian counterpart $\equiv 1$.
Wherever possible, we have compared our results
with the earlier $SU(2)$ study and PCM.
We also display the distribution function(s)
in the color space, which none 
of the other models can yield, be they
flux tube based or pQCD based. All the results will
be shown for the quarks and the gluons separately.
We mention here that it
is the color Flux-tube model within which one 
obtains  the evolution
of the mean background chromoelectric field, 
which is absent, either
in HIJING or PCM. The evolution of such a background field
has a greater impact on the acceleration of the partons present
in the system, and hence on the observed signatures.
For example the c$\bar{c}$ pair, which is
produced in early collisions is acted
by this back ground field along with $c\bar{c}$ potential, to
evolve into a physical $J/{\psi}$.

\subsection{Comparison with SU(2) results}
In the earlier work \cite{nayak}
we had argued that the assumption of the
so called abelian dominance which was made in a large number
of transport studies \cite{baym,kajantie,banerjee}
does not receive any justification
from a proper study of a non-abelian
transport equation. The present
study reinforces the same idea, as it is only
to be expected. In 
particular, note that a determination of quantities such as the
vacuum current or the polarization current 
\cite{eskola} will be 
particularly suspect in view of the fact that 
the effective charge
will only be a fraction of the true charge 
(see below). There now
arises the question of the dependence of the results on $N$ if
one employs the gauge group SU(N) in solving Equation (3).
Although we are not in a position to 
make a strict comparison between SU(2)
and SU(3) in this paper because the SU(2) study had ignored the
gluonic terms altogether, it is still useful 
to see what a limited
comparison can yield.
We shall restrict ourselves to the value $\tau_c = 0.2 fm$ to
contrast very briefly the results of 
SU(2) \cite{nayak} and SU(3).

We shall consider the time dependence of the electric field,
the energy density and the number density, shown in Figs. 1-3.
First of all, it may be noted that the SU(3) quantities
evolve much more rapidly than their SU(2) counterparts.
In fact, while the number density and the 
energy density have attained
their maximum value around $1 fm$, 
at which time the field also
has considerably decayed, there is hardly 
any activity in the SU(2) case even upto 1.5 $fm$. Indeed, 
at $\tau \sim 1 fm$, the number density for SU(3)
is larger by a factor $\sim 10$, and the energy, by a 
factor $\sim 5$, even if we consider only the quark sector.
Please note that it is not that the energy is merely
apportioned
between quarks and gluons; the existence of a
second channel has in
no way decreased the flow to the quark sector.

More significant is the fact that it is the SU(3) flow that
shows the required trend towards a hydrodynamic flow. The SU(2)
coumterparts fail to show any such trend all that way upto
1.5 $fm$. We may certainly expect a hydrodynamic flow
to occur at much later times, but in all likelihood it does not
seem to happen at any realistic value of $\tau$.
We shall discuss
the SU(3) flow in more detail in the next section. On the
whole, the plasma is denser for SU(3) due to increased
volume in phase space. 

\subsection{Discussion of the results}
We now discuss the results for SU(3) in some detail, for three
values of $\tau_c$ ~-~ 0.001, 0.2 and 5 $fm$. 
The first (last) choice
corresponds to the instantaneous hydrodynamic (collisionless)
case, in the time scale set by the initial energy density.
The quark and the gluonic contributions will be
shown separately.
It is clear from Figs. 4 and 5 that the quarks and the gluons
show the same behaviour regarding the energy density and the
number density as functions of proper time. 
The quark contribution is larger, partly because
we have considered two flavors, although the source term 
\cite{gyulassy} favors
a larger rate for gluons in the color space. The situation will
probably get reversed if we include the more correct
perturbative source term for quark \cite{bhal2} and gluon
production \cite{fnote2}. In that case,
the single and three gluon
production rates are of the same order as that of two gluons.
That the two sectors will continue to show the
same trend may be reliably assumed.
The important common feature that Figs. 1, 4 and 5 show is the
nature of the 
evolution at times later than 1 $fm$. The curves suggest
an approach to the hydrodynamic flow. Note that a hydrodynamic
flow would imply that $n(\tau), \epsilon(\tau)$ and $T(\tau)$
behave like $\tau^{-1}, \tau^{-4/3}$ and $\tau^{-1/3}$
respectively. We find that the corresponding exponents 
are $-0.7, -1.23$ and $-0.31$ respectively. While the
temperature scaling suggests a close approach to the free flow regime,
the other two exponents show the
presence of drag \cite{fnote3}, implying
that collisions have not completely ceased. It may be expected
that full hydrodynamic flow will take over around $2 fm$.

First of all let us compare the results for quarks and gluons
separately, for $\tau_c= 0.2 fm$
before presenting the results for different values
of $\tau_c$. In fig-4 we have presented the scaled particle
energy densites. In contrast to 
HIJING and PCM, we obtain  lower values for energy and
number densities for gluons than that of quarks and antiquarks
together (fig-5).
For a complete study, a source term for hard parton production 
is also required
and this term can be obtained from minijet
production at these collider energies, following say ref
\cite{eskola}.
The importance of the non-perturbative contribution may be
gauged qualitatively by noting that the peak number density
for quark-antiquark together(for two flavors)
in this model is $\simeq$ 60 /$fm^3$ in contrast to the PCM
value of $\simeq$ 120 /$fm^3$ (for three massless flavors). 
The energy density for quark sectors in PCM is also larger by
the same factor, implying that the average energy per particle
is of the same order in both the cases.

Let us now consider $<cos^2\theta>$ (where
$\theta$ is the angle between color charge and the 
chromoelectric field in color space).
This quantity is a good bench mark 
to characterise the `non-abelan'ness of the system, and we 
present our results in fig-6.
This quantity is gauge invariant and physical.
It may be seen that this value
saturates at $\theta \simeq \pi/4$, for both quarks and gluons.
This value was always
unity in an abelian plasma, as $\theta \equiv 0$ in that case. 

This has a direct impact on the equilibration of the plasma.
The equilibration is faster around $\theta \simeq \pi/2$
where the background field effect is zero, and is slower at
$\theta \simeq 0$. This is understood as follows. At larger
value of $\theta$, say around $\pi$/2, there is no acceleration
of partons by the background electric field. Only
collisions are present at this angle, and hence the rate
of equilibration is faster. On the otherhand when the angle
$\theta$ = 0, the acceleration of partons by background 
field retards the equilibration of the 
partons.
At any intermediate values of $\theta$, the rate of 
equilibration lies in between these two extreme values.
So this average value, which is purely
due to the non-abelian nature of the theory, has a major role
in the equilibration of the plasma. This is more clearly shown
in the distribution functions which carry the color degrees 
of freedom explicitly(see fig-7 and fig-8).

We conclude the discussion at $\tau_c=0.2 fm$ by making an interesting
observation. As can be seen in fig-6, the value of $<cos^2 \theta>$ 
fluctutates around its mean value for both quarks and gluons, with
similar fluctuations in other quantities as well - although it is not
that pronounced in them. While it certainly indicates 
that the distribution
is approaching its equilibrium value asymptotically, 
the question is whether
the fluctuations would persist even as the system hadronises. 
Unfortunately,
it is difficult to make any definite assertion at this stage. Indeed, as
pointed earlier, evolution of the system beyond $3-4 fm$ requires a full
treatment beyond the assumption of the Bjorken scenario in the central
region. It is also not known when the hadronization exactly sets in.
However, if one assumes that the fluctuations do indeed persist, it
may quite well happen that it will manifest as an anisotropy in
the parton momentum distribution. This has been pointed out in a
study of the related `color filamentation' by 
Mrowczynski \cite{stanislaw}.
It is also possible \cite{fnote4}, 
although we do not know in what manner, that these
fluctuations in the hadronic phase show up as disoriented chiral
condensates. It is only a more complete and rigorous study that can
settle the status of these speculations.

Now we present the data for the other two values of $\tau_c$,
namely $\tau_c = 0.001 fm$ and $\tau_c = 5 fm$, corresponding
to hydrodynamic and collisionless limits respectively.
These results are then compared with
$\tau_c = 0.2 fm$. Consider first the number density $n(\tau)$.
This quantity is seen to 
be sensitive to the values of $\tau_c$.
As can be seen from fig-9, $n(\tau)$ is larger in
the hydrodynamic limit, where the maximum value is around 120
per $fm^3$, at $\tau \simeq 1.0 fm$. As $\tau_c$ is increased 
to 0.2 $fm$ we find a lesser value, the
maximum value being around 80/$fm^3$. This trend
continues for collosionless limit where 
particle density is still less. The increase in number density
as $\tau_c$ is decreased is expected in the context of
classical theory. This is because, 
the collision time in any classical non-equilibrium theory
depends roughly on the inverse of the number density, apart
from the other factors. The behaviour of number densities 
in fig-9 for different values of $\tau_c$ in our calculation 
also reflects the above fact.

In fig-10 we present
the scaled energy densities for different values of $\tau_c$.
Unlike the number density, there is no strong dependence 
on $\tau_c$.
This means that the average energy per particle is different
for different values of
$\tau_c$. We find that for $\tau_c = 0.001 fm$,
the maximum average energy per
particle is around 1.0 GeV, whereas it is around 2.0 GeV for
$\tau_c=0.2 fm$ and 4.0 GeV for $\tau_c= 5.0fm$.
Note that these high energy
deconfined partons can produce secondary
partons, such as strange quarks and also can break a fully
formed $J/{\psi}$ as analysed by short-distance QCD in 
reference \cite{kha}.

Let us consider the electric field.
It may be seen from fig-11 that 
the decay of the 
field is very slow
at $\tau_c = 0.001 fm$, compared to $\tau_c = 0.2$ and 5 $fm$.
Only 30 percent of the field has decayed in the hydrodynamic
limit in comparison to collisionless and intermediate limits,
where the decay is very large. At the other end in the 
collisionsless limit, the
decay is at a slightly 
slower rate in comparison to $\tau_c$ = 0.2 $fm$,
implying that the maximal conversion
rate is for $\tau_c$ around 0.2 $fm$.
As mentioned earlier, one peculiarity of 
the model is that the field energy is continuously created
due to the recession of the plates even as the field itself
decays to produce particles. To study this, in
fig-12 we present field energy
per unit transverse area as a function of {\it{ordinary}} 
time. For $\tau_c$
= 0.2 $fm$ this energy is much less than that at $\tau_c$ = 5 $fm$
and 0.001 $fm$. This demonstrates that the conversion is indeed 
dominant
at $\tau_c$ = 0.2 $fm$. This is more clearly displayed 
in fig-13, where 
we have plotted the ratio of particle energy per unit 
transverse area to field energy per unit transverse area.

In fig-14 we present the evolution of temperature at 
$\tau_c$ = 0.2 $fm$. The maximum temperature we obtain
in the color flux-tube model is around 300 MeV.
Finally, a brief comment on the 
distribution functions for quarks and
gluons with explicit color dependence (see
fig-7 and 8).
As discussed earlier, it may be seen that 
the equilibration is fastest for
$\theta = \pi/2$, where there is no background effect.
At $\theta = \pi/2$ the thermal equilibration
occurs at a common 
time $\tau \simeq$ 1.0 $fm$ for both quarks and gluons.

\section{conclusion}
 We have studied the production and the
 equilibration of a genuinely 
 non-Abelian plasma with the color degree
 of freedom incorporated in
 both the source term and the
 background term in the transport equation.
 With the realistic gauge group SU(3) 
 that we have considerd here,
 the distribution functions for quarks and gluons 
 get defined in the
 extended phase space of dimension 14. We have added
 the gluonic component into the color 
 flux-tube model, which was so far absent. 

The role of the color degree and that of gluons was found to 
be substantial, and non-trivial. The plasma is denser than 
the SU(2) counter part but is comparable to the Abelian case.
However, it is much cooler than the 
abelian plasma, where the corresponding temperature is 
$\sim 800 MeV$. The value of the effective
charge is larger than that of its counterpart in
SU(2), contrary to naive expectations.
The peak energy per particle is $\simeq$ 2 GeV which indeed 
is the demarcating scale \cite{hijing} between soft and hard 
processes. In short, the non-perturbative aspects of the 
evolution of QGP offer a rich variety of results which will have to 
be combined with the perturbative studies in order to abtain
a complete description of the production and evolution of
quark-gluon plasma.

        \newpage

\subsection*{Figure captions}
\noindent
{\bf FIG.~1.} Decay of the chromoelectric field as a function
of proper time (in units of fermi), for $\tau_c=.2 fm$. 
The solid line refers to SU(3)
case, and the dashed line to SU(2) case.

\noindent
{\bf FIG.~2.} The particle energy density scaled w.r.t the
initial field energy density as a
function of proper time (in units of fermi), for 
$\tau_c=.2 fm$. The solid line refers to SU(3)
case, and the dashed line to SU(2) case.

\noindent
{\bf FIG.~3.} The particle number density as a
function of proper time (in units of fermi), for 
$\tau_c=.2 fm$. The solid line refers to SU(3)
case, and the dashed line to SU(2) case.

\noindent
{\bf FIG.~4.} The particle energy density scaled w.r.t the
initial field energy density as a
function of proper time (in units of fermi), for 
$\tau_c=.2 fm$. The solid line refers to total
energy density, upper dashed line to quark plus antiquark
energy density, and lower dashed line to 
the gluon energy density.

\noindent
{\bf FIG.~5.} The particle number density 
as a
function of proper time (in units of fermi), for 
$\tau_c=.2 fm$. The solid line refers
to total number density, upper
dashed line to quark plus antiquark
number density, and lower dashed line to 
the gluon number density.

\noindent
{\bf FIG.~6.} $<cos^2\theta>$ as a function of proper time
(in fermi) at $\tau_c=.2 fm$ for quark (solid line) and gluon 
(dashed line)

\noindent
{\bf FIG.~7.}  $f_q/f_q^{eq}$  
as a function of proper time at $p_t=300 MeV$, 
$\xi=0$ and $\tau_c=0.2 fm$, for different values of
$\theta$. Solid line corresponds to $\theta = \pi/2$.

\noindent
{\bf FIG.~8.}  $f_g/f_g^{eq}$  
as a function of proper time at $p_t=300 MeV$, 
$\xi=0$ and $\tau_c=0.2 fm$, for different values of
$\theta$. Solid line corresponds to $\theta = \pi/2$.

\noindent
{\bf FIG.~9.} The particle number density as a
function of proper time (in units of fermi), for 
$\tau_c=.2 fm$ (solid line), for $\tau_c =
0.001 fm$ (upper dashed line) and for $\tau_c = 5 fm$
(lower dashed line).

\noindent
{\bf FIG.~10.} The particle energy density scaled w.r.t the
initial field energy density as a
function of proper time (in units of fermi), for 
$\tau_c=.2 fm$ (solid line),
for $\tau_c = 0.001 fm$ (upper dashed line)
and for $\tau_c = 5 fm $ (lower dashed line).

\noindent
{\bf FIG.~11.} Decay of the chromoelectric field as a function
of proper time (in units of fermi), for 
$\tau_c=.2 fm$ (solid line),
for $\tau_c = .001 fm$ (upper dashed line) and for 
$\tau_c = 5 fm$ (lower dashed line).

\noindent
{\bf FIG.~12.} The field energy/unit transverse area(in units
of GeV/{$fm^2$}) for $\tau_c$
= 0.2 $fm$ (solid line), for 
$\tau_c$ = 0.001 $fm$ (upper dashed line)
and for $\tau_c$ = 5 $fm$ (lower dashed line),
as a function of ordinary 
time (in fermi).

\noindent
{\bf FIG.~13.} The ratio of particle energy per unit transverse
area to field energy per unit transverse area
for $\tau_c$
= 0.2 $fm$ (solid line), for $\tau_c$ = 5 $fm$ (upper dashed line)
and for $\tau_c$ = .001 $fm$ (lower dashed line),
as a function of ordinary time (in fermi).

\noindent
{\bf FIG.~14.} Evolution of temperature as a function of
proper time (in fermi), for $\tau_c$ = 0.2 $fm$.


\begin{thebibliography}{99}
\bibitem{fnote1}  Increasing the accuracy by another 5 percent
       does not substantially alter the estimate.
\bibitem{fnote3} see ref \cite{eskola}.
       The Ohmic heating in the reference is 
       realised by the particle production in our model.
\bibitem{fnote4} see, e.g., K. Rajagopal, CALT-68-2104, 
       who has discussed the possible impact of non-equilibrium
       state on the formation of disoriented chiral
       condensates.
\end{thebibliography}
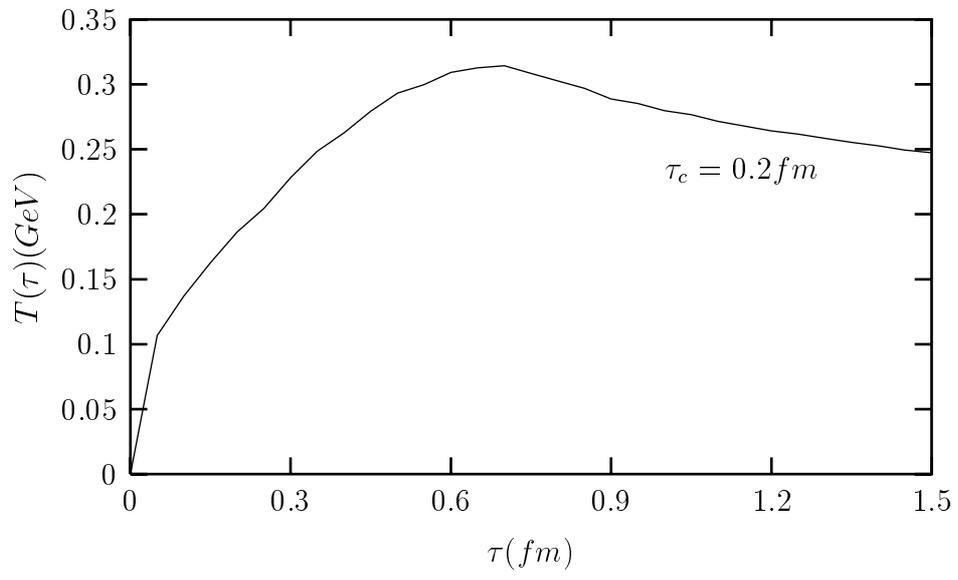
\end{document}